\documentclass
[showpacs,prl,amsfonts,amssymb,oneside,preprint,balancelastpage,letterpaper,onecolumn,12pt,titlepage]{revtex4}%
\usepackage{amsfonts}
\usepackage{amsmath}
\usepackage{amssymb}
\usepackage{graphicx}%
\begin{document}
\title{Phase--locking of a Nonlinear Optical Cavity via Rocking: Transmuting Vortices
into Phase Patterns}
\author{Adolfo Esteban--Mart\'{\i}n, Manuel Mart\'{\i}nez--Quesada, Victor B.
Taranenko, Eugenio Rold\'{a}n, and Germ\'{a}n J. de Valc\'{a}rcel}
\affiliation{Departament d'\`{O}ptica, Universitat de Val\`{e}ncia, Dr. Moliner 50,
46100--Burjassot, Spain}

\begin{abstract}
We report experimental observation of the conversion of a phase--invariant
nonlinear system into a phase--locked one via the mechanism of rocking [de
Valc\'{a}rcel and Staliunas, Phys. Rev. E \textbf{67}, 026604 (2003)]. This
conversion results in that vortices of the phase--invariant system are being
replaced by phase patterns such as domain walls. The experiment is carried out
on a photorefractive oscillator in a two-wave mixing configuration. A model
for the experimental device is given that reproduces the observed behavior.

\end{abstract}

\maketitle

\textit{Introduction}.--The symmetry properties of the order parameter phase
in extended nonlinear systems determine the type of localized structures
supported by these systems. On the one hand phase-invariant systems display
vortices, which are phase defects of the order parameter, around which the
phase changes in $2\pi$, forcing the order parameter to be null at the vortex
center. Self-oscillatory systems represent a paradigm of such behavior as the
oscillating state reached after a homogeneous Hopf bifurcation can have any
phase (the system is autonomous). Examples of such systems are certain
chemical reaction (like the Belousov-Zhabotinsky reaction) and several optical
systems (like lasers, nondegenerate optical parametric oscillators, OPOs, and
nondegenerate wave mixing photorefractive oscillators). On the other hand
there are nonlinear systems with broken phase invariance that, obviously,
cannot support vortices. Among them, especially interesting systems are those
displaying phase bistability. Such systems support a different type of
localized structure, the domain wall, which connects spatial regions where the
order parameter passes from one homogeneous state, $F_{0}$, to the equivalent,
symmetric state, $-F_{0}$ \cite{Cross93} so that the phase changes by $\pi$
from one side of the wall to the other one. A paradigm of such type of
phase-bistable system is the degenerate OPO.

The question we address here is: Is it possible to convert a phase-invariant
system into a phase-bistable one by some simple external action? In other
words, can a system displaying vortices be forced to display domain walls? An
old and well known answer to these questions consists in the periodic forcing
of a self-oscillatory system at a frequency around two times its natural
oscillation frequency $\omega_{0}$ (resonance $2:1$). Under that type of (so
called) parametric driving the originally phase-invariant system is predicted
\cite{Coullet90} and experimentally observed \cite{BZ1,BZ2} to transform into
a phase-bistable system, exhibiting domain walls. But while parametric driving
is useful in many contexts it is not so in general in nonlinear optics. For
instance, a laser emitting at frequency $\omega_{0}$ is insensitive to a
forcing at $2\omega_{0}$ because the gain line is extremely narrow as compared
with the magnitude of the optical frequency $\omega_{0}$. Yet, a new mechanism
coined \textit{rocking} \cite{deValcarcel03} has been recently proposed to
overcome such limitation of the usual parametric driving. Rocking consists in
forcing a self-oscillatory system around its natural oscillation frequency
(resonance $1:1$) so that the forcing amplitude is periodically modulated in
time at a frequency $\omega\ll\omega_{0}$. Thus rocking is a multifrequency
forcing (in its simplest case, a sinusoidal modulation of the forcing
amplitude, it is a bichromatic forcing) around the resonance $1:1$ of a
self-oscillatory system. We want to remark that unlike the parametric driving,
any self-oscillatory system (including lasers and any nonlinear optical
system) should be sensitive to rocking as this forcing acts on the main
resonance of the system.

In order to gain insight into the rocking idea, consider a particle in a 2D
space with coordinates $\left(  x,y\right)  $ under the action of a potential
with the sape of a sombrero --having a maximum at the origin $\left(
0,0\right)  $ and a degenerated minimum at $x^{2}+y^{2}=r_{\min}^{2}$. This is
a phase-invariant system, what means that the equilibrium position is
phase(angle)-degenerated. Suppose now that this potential is rocked along the
$x$-direction (e.g., by adding a term $x\sin\omega t$ to the original
potential). In this case the phase degeneracy is obviously broken and now the
system tends to be around $\left(  x=0,y=\pm r_{\min}\right)  $: The system is
now phase-bistable. This pictorial image illustrates the physical rationale
behind the rocking idea: The coordinates of the fictious particle correspond
to the real and imaginary parts of the complex amplitude of oscillations of
the self-oscillatory system, whose evolution derives from a sombrero-like
potential in the simplest case, as originally introduced in
\cite{deValcarcel03} through an analysis of a complex Ginzburg-Landau
equation, which is the simplest model for spatially extended self-oscillatory
systems \cite{Cross93}. As shown in \cite{deValcarcel03}, under the action of
rocking the order parameter of the system $F\left(  \mathbf{r},t\right)  $
(e.g., the laser electric field complex amplitude) develops two components,
$F\left(  \mathbf{r},t\right)  =F_{\omega}\left(  t\right)  +i\psi\left(
\mathbf{r},t\right)  $, where $F_{\omega}$ is a $\frac{2\pi}{\omega}%
$--periodic function of time that follows the modulation, at frequency
$\omega$, of the forcing amplitude and $\psi$ is a phase-bistable, reduced
order parameter that can display domain walls, phase domains, and phase domain
solitons \cite{deValcarcel03}. Even if the original prediction is based on a
complex Ginzburg-Landau model, rocking is expected to be of wide applicability
as the phase-bistability mechanism introduced above seems to be quite model independent.

In this Letter we give the first experimental evidence of rocking induced
phase-bistability and the associated formation of phase domains and domain
walls. Our system is a (nondegenerate) two-wave mixing photorefractive
oscillator (PRO), which highly resembles a laser from the nonlinear dynamics
viewpoint. Rocking is done by injecting an amplitude modulated laser beam into
the resonator. We show that in the absence of rocking (free running
configuration) the PRO exhibits vortices, which are converted into phase
domains under the action of rocking. Experiments performed under a quasi 1D
transverse geometry (in order to avoid curvature effects) show that the system
displays domain walls in a window of rocking modulation frequencies, in
agreement with \cite{deValcarcel03}. A model for the experimental device is
presented that reproduces the experimental findings. In a limit, such model
reduces to a parametrically driven complex Swift-Hohenberg equation, which
represents a generalization of \cite{deValcarcel03} to other types of order
parameter equations and hence enlarges the range of applicability of rocking.

\textit{Experimental setup.}-- The PRO, Fig. 1, consists of a photorefractive
\textrm{BaTiO}$_{3}$ crystal placed inside a (near) self-imaging ring
resonator \cite{Arnaud69} of cavity length $1.2\mathrm{m}$ (the resonator
free-spectral range is $250$\textrm{MHz}), similar to that used in
\cite{Vaupel95}. The effective cavity length is approximately $2$\textrm{cm},
which is actively stabilized by means of piezomirror PZT1 in order to have a
precise control on cavity detuning (the difference between the frequencies of
the pump field and that of the closest cavity mode). The crystal is pumped by
a singlemode $514$\textrm{nm} \textrm{Ar}$^{+}$ laser with a power around
$100$\textrm{mW cm}$^{-2}$. An amplitude modulated beam (the rocking beam)
coherent to the pump field is injected into the cavity. The c+ crystal axis is
oriented so that gain is maximized when all three beams (pump, oscillation and
injection) are extraordinarily polarized. The intracavity slit D is placed in
a Fourier plane in order to make the system quasi 1D in the transverse
dimension \cite{Esteban04}. Finally, there is the possibility of injecting a
tilted coherent beam in order to "write" domain walls as in
\cite{Esteban05,Esteban05b}.

The rocking beam must be an amplitude modulated field of zero mean: We chose
to use a field of constant amplitude, whose phase changes exactly by $\pi$ in
every half period. We do this by injecting into the cavity a beam coming from
the pumping laser, after being reflected on piezomirror PZT2, Fig. 1, which is
moved periodically back and forth by half a wavelength (the modulation
frequency is on the order of $1$\textrm{Hz}). This way the rocking has a pure
amplitude modulation (only the sign of the field amplitude changes). This
operation must be very precise as phase jumps different to $\pi$ do not
produce the desired result.

\textit{Experimental results.--} The cavity length is chosen for the system be
in (almost) exact resonance. This is not mandatory but is the simplest way for
the rocking field be resonant with the intracavity field (see below the model
section). When the rocking beam is off and the intracavity slit is open, the
system spontaneously forms vortices in the output field, as expected
\cite{Vortices}. Under these conditions, the application of a rocking beam is
able to transform vortices into phase domains, as shown in Fig. 2. This result
is a direct demonstration of the rocking induced phase-bistability.

Two dimensional phase domains are transient structures due to curvature
\cite{deValcarcel03}. Hence we performed a series of experiments under quasi
1D conditions (by narrowing the intracavity slit), which allow stable domain
walls in phase bistable systems \cite{Esteban05}. These domain walls are
however unstable in 1D phase invariant systems: If a domain wall is injected
into the free running PRO, Fig. 3(a), we observe that the domain wall vanishes
with time, Fig. 3(b), and is replaced by a spatially uniform state, Fig. 3(c).
On the contrary, if the PRO is rocked completely different results are
obtained. Now an injected domain wall, Fig. 3(e), remains stable and fixed,
Figs. 3(f) and (g). (The amplitude modulation frequency of the rocking beam
was $1.5$\textrm{Hz} and its intensity was of the same order as that of the
output field in the absence of rocking.) Figure 3(h) evidences that we are in
the presence of an Ising wall \cite{Coullet90,Esteban05}. The structure is
robust and this proofs that the system is now phase bistable. The ability of
rocking to sustain domain walls is thus proven.

There remains to clarify which is the effect on the performance of the system
of changing the modulation frequency and amplitude of the rocking beam. We
have observed that the frequency of modulation cannot be too small or too
large: For modulation frequencies of $0.1\mathrm{Hz}$ or below rocking cannot
sustain the injected domain wall. The same happens for modulation frequencies
greater than $10\mathrm{Hz}$. It seems that modulations from $1$ to
$3\mathrm{Hz}$ are optimal, which are on the order of the inverse of the
photorefractive grating decay time. While we do not have definite
measurements, we can say that rocking is effective in a window of rocking
intensities as well.

All previous features are in agreement with the predictions of Ref.
\cite{deValcarcel03}. The theory is however based on a Ginzburg-Landau model,
which should not be valid near resonance \cite{LMN,OPO1,OPO2}, the case
considered here. In order to justify theoretically the results on a more firm
basis we thus proceed to model the PRO.

\textit{Theory.--} We adopt the two-wave mixing PRO model for a purely
diffusive photorefractive crystal \cite{Bortolozzo}, such as \textrm{BaTiO}%
$_{3}$, generalized to account for the injection of the rocking beam (details
will be given elsewhere). The model equations, suitably normalized, can be
written as:%

\begin{gather}
\sigma^{-1}\partial_{t}F=-\left(  1+i\Delta\right)  F+ia\nabla^{2}%
F+N+F_{\mathrm{R}}\left(  t\right)  ,\label{f1}\\
\partial_{t}N=-N+g\frac{F}{1+\left\vert F\right\vert ^{2}}, \label{g1}%
\end{gather}
where $F\left(  \mathbf{r},t\right)  $ is the slowly varying envelope of the
intracavity field, $N\left(  \mathbf{r},t\right)  $ is the complex amplitude
of the photorefractive nonlinear grating, $\mathbf{r}=\left(  x,y\right)  $
are the transverse coordinates, $\nabla^{2}=\partial_{x}^{2}+\partial_{y}^{2}%
$, $\sigma=\kappa\tau$ is the product of the cavity linewidth $\kappa$ with
the photorefractive response time $\tau$ ($\sigma\gtrsim10^{8}$ under typical
conditions, and $\tau\sim1\mathrm{s}$), $t$ is time measured in units of
$\tau$, the detuning $\Delta=\left(  \omega_{\mathrm{C}}-\omega_{\mathrm{P}%
}\right)  /\kappa$ ($\omega_{\mathrm{P}}$ and $\omega_{\mathrm{C}}$ are the
frequencies of the pump and of its nearest cavity longitudinal mode,
respectively), $a$ is the diffraction coefficient (that depends upon geometry
\cite{Esteban04} and can take either sign), $F_{\mathrm{R}}$ is the complex
envelope of the rocking (injected) field, and $g$ is the (real) gain parameter
that depends on crystal parameters and on the geometry of the interaction. The
actual intracavity field, $\mathcal{E}$, and rocking field, $\mathcal{E}%
_{\mathrm{R}}$, read $\mathcal{E}=\operatorname{Re}FE_{\mathrm{P}}%
e^{-i\omega_{\mathrm{P}}\tau t}$ and $\mathcal{E}_{\mathrm{R}}%
=\operatorname{Re}F_{\mathrm{R}}E_{\mathrm{P}}e^{-i\omega_{\mathrm{P}}\tau t}%
$, respectively, where $E_{\mathrm{P}}$ and $\omega_{\mathrm{P}}$ are the
complex amplitude and the angular frequency of the pumping laser field. Note
that $E_{\mathrm{P}}$ acts just as a scaling factor (it is unrelated with the
gain parameter $g$) \cite{Bortolozzo,Yeh93}.

For $F_{\mathrm{R}}=0$ the model is equivalent to that in \cite{Bortolozzo},
which holds the continuous symmetry $\left(  F,N\right)  \rightarrow\left(
Fe^{i\phi},Ne^{i\phi}\right)  $. Hence the system is phase invariant in the
absence of forcing. This free running PRO model has two main solutions
\cite{Bortolozzo}: The trivial solution $F=N=0$, and the family of
traveling-wave solutions (parametrized by the wavevector $\mathbf{k}$)
$F=\sqrt{g/g_{0}-1}e^{i\left(  \mathbf{k}\cdot\mathbf{r}-\Omega t\right)  }$
and $N=\left(  1+i\Omega\right)  F$, with $g_{0}=1+\Omega^{2}$,$\;$and
$\Omega=\frac{\sigma}{\sigma+1}\left(  \Delta+ak^{2}\right)  \rightarrow
\Delta+ak^{2}$ as $\sigma\gg1$. One easily sees that: (i) for $\Delta/a>0$ the
oscillation threshold is minimum for $k=0$ (on--axis emission), occurs at
$g=1+\Delta^{2}$, and the frequency of the generated field is shifted by
$\Omega=\Delta$ from that of the pump beam; and (ii) for $\Delta/a<0$ the
threshold is minimum for $k=\sqrt{-\Delta/a}$ (off--axis emission), occurs at
$g=1$, and there is no frequency shift ($\Omega=0$). We checked these features
in our experiment. In particular cavity resonance was determined by
interfering the emitted field with a reference coming from the pumping laser:
The cavity length at which the beating frequency ($\Omega$) passes from zero
to a non null value (or vice versa) corresponds to exact resonance
\cite{Bortolozzo}. These facts, together with the phase invariance of the
model, make the PRO in the two-wave mixing configuration a system largely
equivalent to a large aspect ratio laser \cite{Bortolozzo,Staliunas}.

In the following we use $F_{\mathrm{R}}=R\cos\omega t$ ($R$ real without loss
of generality), which is the simplest form of amplitude modulation and
corresponds to a bichromatic injected signal. In order to give analytical
evidence of the rocking induced phase-bistability in the PRO model we perform
an asymptotic expansion of Eqs. (\ref{f1},\ref{g1}) based on the of multiple
scales technique \cite{Nayfeh}. In order to approach the experimental
conditions we assume small detuning, and large cavity linewidth: We take
$\Delta=O\left(  \varepsilon\right)  $ , $\sigma=O\left(  \varepsilon
^{-4}\right)  $ with $\varepsilon$ a smallness parameter ($0<\varepsilon\ll1$)
(the final result does not depend on the precise scaling for $\sigma$,
whenever $\sigma\gg1$ is assumed). We further assume that gain is close to
threshold, $g=1+O\left(  \varepsilon^{2}\right)  $, and that the rocking
parameters verify $R=O\left(  \varepsilon\right)  $, $\omega=O\left(
1\right)  $. The analysis, similar to that performed in laser \cite{LMN} and
nondegenerate OPO \cite{OPO1,OPO2}\ models, yields $N=\left[  1+i\left(
\Delta-a\nabla^{2}\right)  \right]  F-R\cos\omega t$,%
\begin{equation}
F\left(  \mathbf{r},t\right)  =F_{\omega}\left(  t\right)  +i\psi\left(
\mathbf{r},t\right)  +O\left(  \varepsilon^{3}\right)  , \label{F}%
\end{equation}
$F_{\omega}\left(  t\right)  =\frac{R}{\omega}\left[  \left(  1-2i\Delta
\right)  \sin\omega t+\left(  1-i\Delta\frac{\omega^{2}-1}{\omega^{2}}\right)
\omega\cos\omega t\right]  $, the order parameter $\psi$ verifying a complex
Swift-Hohenberg equation with parametric gain:
\begin{align}
\partial_{t}\psi &  =\gamma\psi^{\ast}+\left(  g-1-2\gamma\right)
\psi-\left\vert \psi\right\vert ^{2}\psi\nonumber\\
&  +i\left(  a\nabla^{2}-\Delta\right)  \psi-\left(  a\nabla^{2}%
-\Delta\right)  ^{2}\psi, \label{CSHE}%
\end{align}
where $\gamma=\frac{R^{2}}{2}\frac{1+\omega^{2}}{\omega^{2}}$ is the rocking
parameter (note that $\gamma\geq\frac{R^{2}}{2}$). In the absence of rocking
($\gamma=0$), Eq. (\ref{CSHE}) is phase invariant and is isomorphic to those
describing lasers \cite{LMN} and nondegenerate OPOs \cite{OPO1,OPO2}, as well
as highly resembles that for drift-type PROs \cite{Staliunas}. The role of
rocking is clearly appreciated in\ Eq. (\ref{CSHE}): It introduces an
phase-sensitive gain (first term in the r.h.s.) that breaks the original phase
invariance of the undriven system down to the discrete one $\psi
\rightarrow-\psi$. Thus the PRO becomes phase bistable. The spatially uniform
steady states of Eq. (\ref{CSHE}) are given by $\psi=\pm\left\vert
\psi\right\vert e^{i\varphi}$, $\left\vert \psi\right\vert ^{2}=\mu
-2\gamma+\sqrt{\gamma^{2}-\Delta^{2}}$, $\mu=g-1-\Delta^{2}$, $e^{2i\varphi
}=\frac{\sqrt{\gamma^{2}-\Delta^{2}}-i\Delta}{\gamma}$ (other two,
intrinsically unstable states exist as well, which we do not consider). These
phase-locked states exist if $\left\vert \Delta\right\vert \leq\gamma
\leq\gamma_{\max}$ ($\gamma_{\max}=\frac{2\mu}{3}+\frac{1}{3}\sqrt{\mu
^{2}-3\Delta^{2}}$) as far as $\mu\geq\sqrt{3}\left\vert \Delta\right\vert $.
These inequalities imply analogous ones for the rocking intensity, $R^{2}$, or
the rocking modulation frequency, $\omega$, depending on which parameter
($\omega$ or $R$) is kept fixed. This prediction supports the experimental
findings described above. Finally note that as $\gamma\geq\frac{R^{2}}{2}$ the
rocked states will exist only if $R^{2}\leq2\gamma_{\max}$, and $\mu\geq
\sqrt{3}\left\vert \Delta\right\vert $, for any $\omega$.

In a series of numerical experiments we have checked the above predictions and
have found that they remain valid even far from the asymptotic limit described
by Eq. (\ref{CSHE}), as it happens in the original proposal
\cite{deValcarcel03}. In order to have additional comparison between theory
and experiment, we present in Fig. 4 some numerical simulations of Eqs.
(\ref{f1},\ref{g1}) for $\sigma=10^{2}$ (large, but not extremely large in
order to avoid stiffness problems), $\Delta=0$, $g=2$ (we estimate that the
gain in the experiment is about 100\% above threshold), $\omega=2\pi$ and
$R=0.5$. Comparison with Fig. 3 shows that results are very similar to the
experimental ones. A message from this theoretical treatment arises: Phase
invariant systems described by order parameter equations of different nature
(like the Ginzburg-Landau and the Swift-Hohenberg complex models) behave
similarly under the influence of rocking.

In conclusion, we have experimentally demonstrated, and theoretically
justified, the ability of the rocking mechanism introduced in
\cite{deValcarcel03} to generate phase-bistable states in otherwise phase
invariant systems. The studied system, a PRO in two-wave mixing configuration,
highly resembles laser systems from the nonlinear dynamics viewpoint. Hence
the results put forward in this Letter should motivate similar experiments in
laser systems, which could have potential applications in the field of
information technologies.

\begin{acknowledgments}
This work was supported by Spanish Ministerio de Educaci\'{o}n y Ciencia and
European Union FEDER through Project FIS2005-07931-C03-01. We gratefully
acknowledge fruitful discussions with Javier Garc\'{\i}a Monreal and Kestutis Staliunas.
\end{acknowledgments}

\begin{center}
{\LARGE Figure captions}
\end{center}

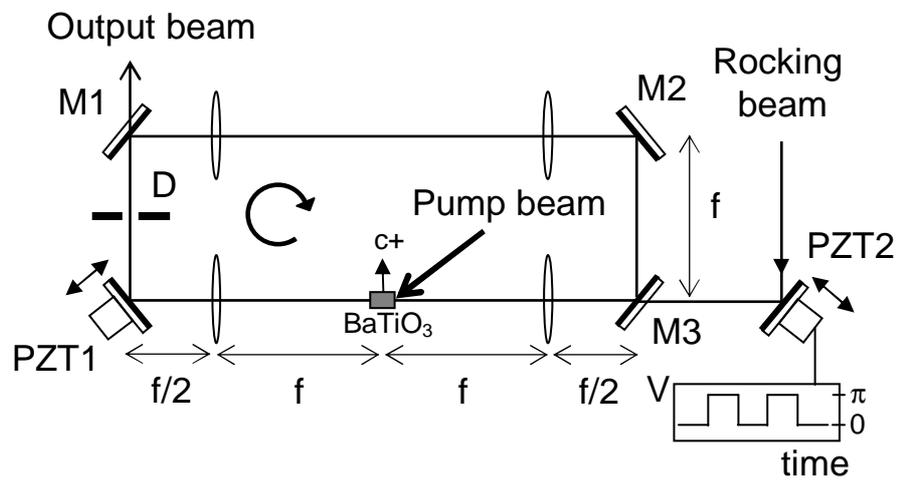
\textbf{Fig. 1.-} Scheme of the experimental setup. A photorefractive
\textrm{BaTiO}$_{3}$ cristal is placed inside the quasi self-imaging resonator
formed by mirrors M1, M2 and M3, piezomirror PZT1 and the four lenses of focal
length $f$. D is a diafragm, and PZT2 is a piezomirror driven by a square-wave
signal, plot $V$ vs. time, that produces the rocking beam.

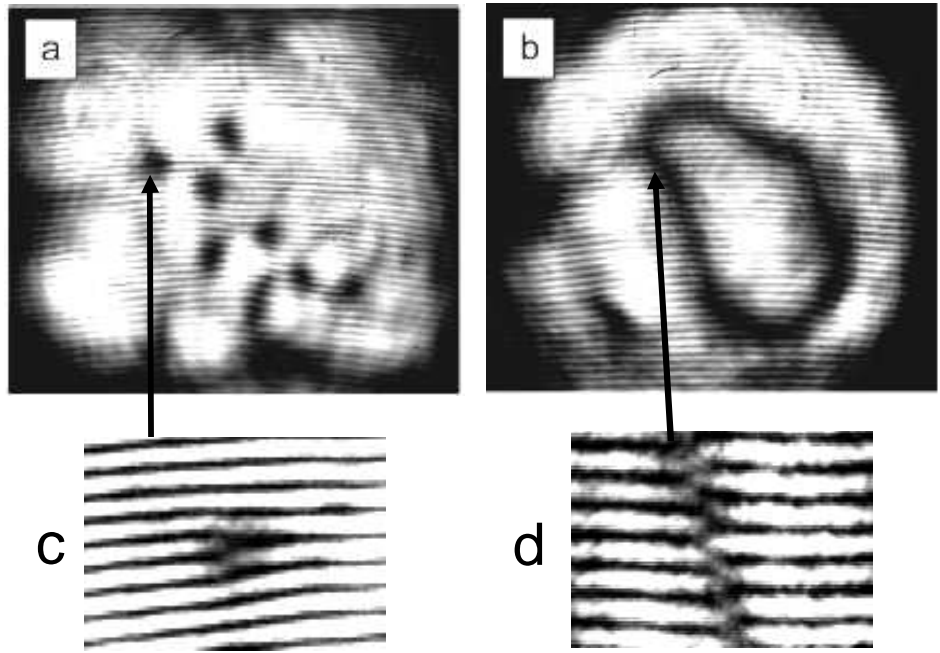
\textbf{Fig. 2.-} Interferometric snapshots of vortices (a) existing in the
free running PRO and phase domain structure (b) generated in the rocked PRO at
resonance.\ Magnifications of the interferometric images in the position of
(c) a vortex showing the annihilation of the interference fringes and (d) of a
part of the domain wall showing the $\pi$ phase jump.

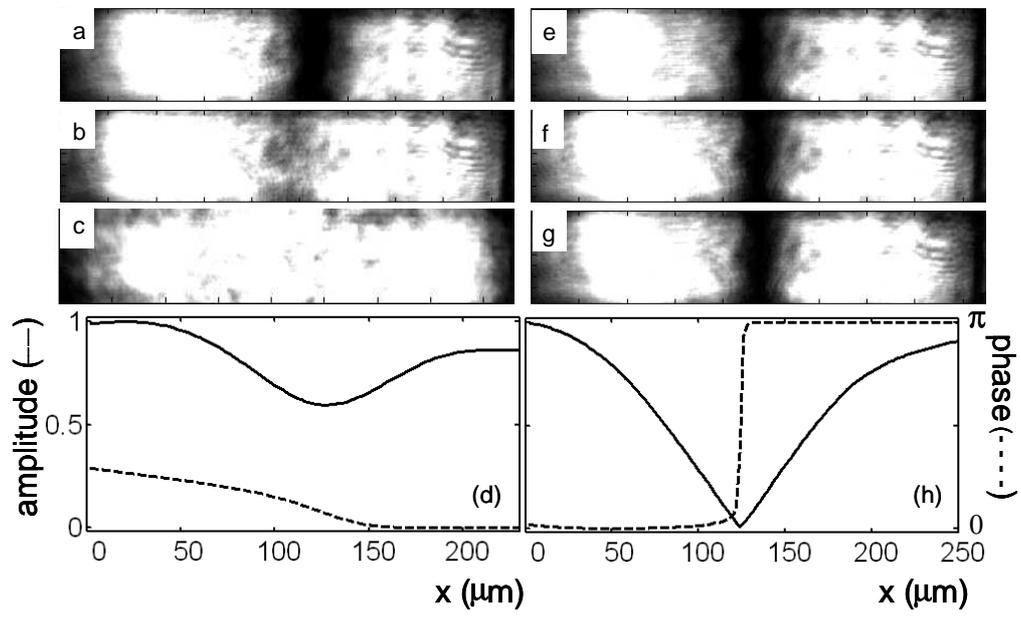
\textbf{Fig. 3.-} Experimental snapshots of injected domain walls (DW) close
to cavity resonance. (a)-(c): Disappearance of a DW in the free running PRO
(the time interval between snapshots is $5$\textrm{s}). (e)-(g): Stabilization
of the DW in the rocked PRO (the rocking frequency is $1.5$\textrm{Hz }and the
time interval between snapshots is $15\mathrm{s}$). (d) and (h): Horizontal
cuts showing the field amplitude (a.u.) and phase corresponding to snapshots
(b) and (f), respectively. The snapshots transverse dimension is
$1.2$\textrm{mm}.

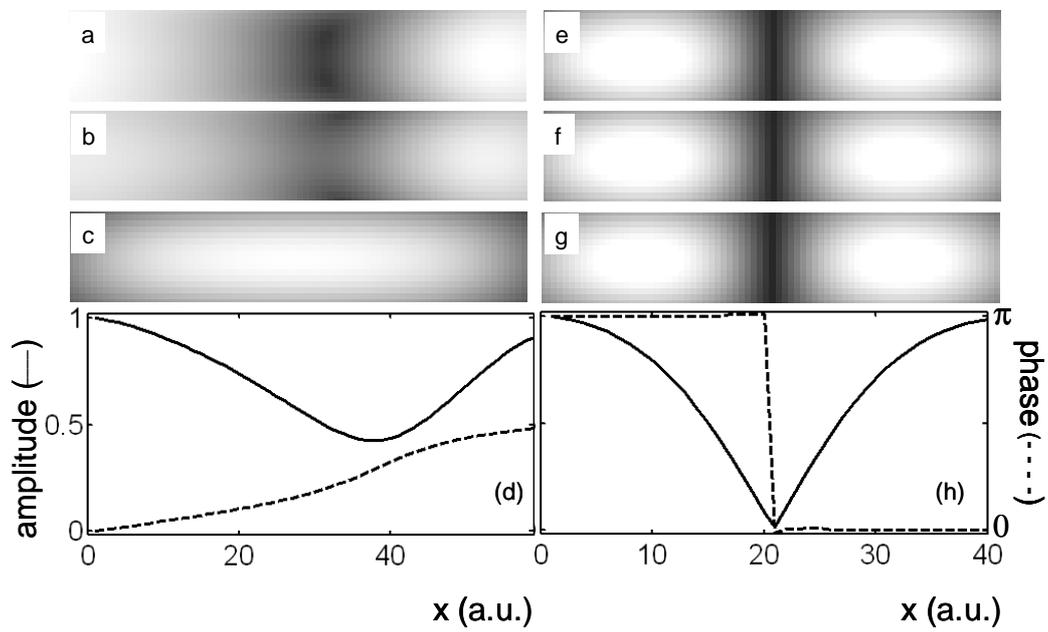
\textbf{Fig. 4.-} As Fig. 3 but from numerical simulations of Eqs. (\ref{f1}),
(\ref{g1}). Parameters: $\sigma=10^{2}$, $\Delta=0$, $g=2$, $\omega=2\pi$, and
$R=0.5$. The length of the horizontal dimension is $21\sqrt{a}$ in (a)-(c) and
(e)-(g), $14\sqrt{a}$ in (d) and $9.3\sqrt{a}$ in (h).


\begin{thebibliography}{99}                                                                                               %


\bibitem {Cross93}M. C. Cross and P. C. Hohenberg, Rev. Mod. Phys.
\textbf{65}, 851 (1993).

\bibitem {Coullet90}P. Coullet, J. Lega, B. Houchmanzadeh and J. Lajzerowicz,
Phys. Rev. Lett. \textbf{65}, 1352 (1990).

\bibitem {BZ1}V. Petrov, Q. Ouyang, and H. L. Swinney, Nature (London)
\textbf{388}, 655 (1997).

\bibitem {BZ2}A. L. Lin, M. Bertran, K. Mart\'{\i}nez, H. L. Swinney, A.
Ardelea, and G. F. Carey, Phys. Rev. Lett. \textbf{84}, 4240 (2000).

\bibitem {deValcarcel03}G. J. de Valc\'{a}rcel and K. Staliunas, Phys. Rev. E
\textbf{67}, 026604 (2003).

\bibitem {Arnaud69}J. A. Arnaud, Appl. Opt. \textbf{8}, 189 (1969).

\bibitem {Vaupel95}M. Vaupel and C. O. Weiss, Phys. Rev. A \textbf{51}, 4078 (1995).

\bibitem {Esteban04}A. Esteban-Mart\'{\i}n, J. Garc\'{\i}a, E. Rold\'{a}n, V.
B. Taranenko, G. J. de Valc\'{a}rcel, and C.O. Weiss, Phys. Rev. A
\textbf{69}, 033816 (2004).

\bibitem {Esteban05}A. Esteban-Mart\'{\i}n, V. B. Taranenko, J. Garc\'{\i}a,
G. J. de Valc\'{a}rcel, and E. Rold\'{a}n, Phys. Rev. Lett. \textbf{94},
223903 (2005).

\bibitem {Esteban05b}A. Esteban-Mart\'{\i}n, V. B. Taranenko, E. Rold\'{a}n,
and G. J. de Valc\'{a}rcel, Opt. Express \textbf{13}, 3631 (2005).

\bibitem {Vortices}K. Staliunas, G. Slekys, and C. O. Weiss, Phys. Rev. Lett.
\textbf{79}, 2658 (1997).

\bibitem {Bortolozzo}U. Bortolozzo, P. Villoresi, and P. L. Ramazza, Phys.
Rev. Lett. \textbf{87}, 274102 (2001).

\bibitem {Yeh93}P. Yeh, \textit{Introduction to Photorefractive Nonlinear
Optics }(John Wiley \& Sons, New York, 1993).

\bibitem {Staliunas}K. Staliunas, M. F. H. Tarroja, G. Slekys, C. O. Weiss,
and L. Dambly, Phys. Rev. A \textbf{51}, 4140 (1995).

\bibitem {Nayfeh}A. H. Nayfeh, \textit{Perturbation Methods} (John Wiley \&
Sons, New York, 2000).

\bibitem {LMN}J. Lega, J. V. Moloney and A. C. Newell, Phys. Rev. Lett.
\textbf{73}, 2978 (1994).

\bibitem {OPO1}S. Longhi and A. Geraci, Phys. Rev. A \textbf{54}, 4581 (1996).

\bibitem {OPO2}V. J. S\'{a}nchez-Morcillo, E. Rold\'{a}n, G. J. de
Valc\'{a}rcel, and K. Staliunas, Phys. Rev. A \textbf{56}, 3237 (1997).

\smallskip
\end{thebibliography}
\end{document}